\documentclass[11pt]{article} 

\usepackage[utf8]{inputenc} 

\usepackage{geometry} 
\usepackage{graphicx} 
\usepackage{mathtools}
\usepackage{amsmath}

\usepackage{amssymb}

\usepackage{array} 
\usepackage{paralist} 
\usepackage{verbatim} 
\usepackage{subfig} 

\usepackage{fancyhdr} 
\usepackage{sectsty}
\allsectionsfont{\sffamily\mdseries\upshape} 

\usepackage[nottoc,notlof,notlot]{tocbibind} 
\usepackage[titles,subfigure]{tocloft} 

\title{A fair monetization model to reconcile authors and consumers of intellectual property}
\author{Evgeny Ivanko\\ \textit{evgeny.ivanko@gmail.com}}

\begin{document}
\maketitle

\section{Introduction}

No one likes to pay for virtual goods. There may be many deep psychological reasons behind it. Maybe we are not happy to pay for intangible stuff, maybe we do not agree with the authorship model, where one single product is copied for no expense and bought for considerable expense many times. Maybe in our fast and dense century we can find so much information around, that it feels ridiculous to pay for it. Anyway, we do not like to pay for intellectual property (IP) (see e.g. \cite{1,2}). 

Intellectual property is a stumbling block and bone of contention for many states, industries, companies and people. In this small article one compromise monetization strategy is proposed, which hopefully may lead to a more satisfactory coexistence of IP manufacturers and consumers. So, how could a manufacturer get his expected profit if we do not like to pay for his job? The motto is ``fair exchange'': you use our IP-product, we use your product (in form of money); when you do not need our product any more, we \textit{change back}. 

In the next section we consider the general structure of the proposed scheme and add a brief discussion. Section 3 is devoted to a simple rough computation answering the question ``how much clients do we need to flourish with this model'' (if you do not like formulas, then skip to the last two paragraphs). The final section gives several brief examples and may prove the most interesting part of this text.

\section{Strategy}

\begin{itemize}
\item Consumer deposits the agreed amount of money for the agreed period of time to the account of manufacturer and get the right to use IP-product during that time
\item Manufacturer may use the invested money at his own discretion, e.g. deposit to a bank, or buy obligations, or invest to development, but he has to return all the money upon the end of the agreement
\item Consumer may wish to retain the IP product after the end of the agreement, then the deal becomes a simple buy-sell
\item Transaction commissions may apply to either the manufacturer or the consumer, but it is better to apply no commissions at all -- the bank managing the manufacturer's account may receive enough profit from permanently large deposit to make transactions free; if the manufacturer is large enough, he may  think of his own transaction-bank
\end{itemize}

The proposed approach blurs the edge between sales of products to customers and sales of shares to investors.
The number of deals in this scheme should be far more than in case of plain buy-sell business, because it seems much easier to make a decision to pay in case you are sure you get your money back. 

The price and duration of contract should be chosen for each business and each product separately and accurately; see several naive examples in Section 4. In particular cases  a manufacturer may even pay a part of his interest to his clients; in this case the edge between IP-business and investment fund starts to thin.

The issues with piracy will surely appear, but probably not greater than in case of usual buy-sell approach. Actually, piracy may even decrease since the proposed approach may seem more fair in the eyes of clients. 

Let us make some very preliminary and approximate computations concerning the effectiveness of the proposed model. If you do not like formulas, you may omit the next section without loosing anything essential.

\section{Brief and crude computation}

How many clients are needed and what should be the deposit size for the seller to flourish? Let us model a bit (however, we do not claim to be realistic). Let $p(x,t)$ be the probability of depositing $x\in\overline{0,M}$ coins for $t\in\overline{0,T}$ days by one average client on an average day, where 

$$\sum_{x=0}^M\sum_{t=0}^T p(x,t)=1.$$

This probability is assumed to be equal for all clients and days. How much money will remain in the account on average? The expected one-client today investment is
\begin{equation}
\label{*}
\tag{$*$}
\sum_{x=1}^M x \sum_{t=1}^T p(x,t).
\end{equation}

For yesterday investments we consider only those that were placed for 2 days or more,
$$\sum_{x=1}^M x \sum_{t=2}^T p(x,t),$$
and so on. Summing over all previous days up to $T$ we have

$$\sum_{k=1}^T\sum_{x=1}^M x \sum_{t=k}^T p(x,t).$$

Now let us choose a simple but plausible function for $p(x,t)$, e.g.

$$p(x,t)=
\begin{dcases*}
\frac{K_1}{x(x+1)}\frac{K_2}{t(t+1)},\quad x\ge 1, t\ge 1;\\
K_0,\qquad x=0\ \text{or}\ t=0.
\end{dcases*}
$$


Please note, it is by no means a function from the industry, it was pulled out of a hat. Taking large $M$ and $T$ (like $M=10^6, T=10^3$) and using the summation formula for telescopic series, the probability of the event ``an average client invests something today'' may be expressed as

$$\sum_{x=1}^M\sum_{t=1}^T \frac{K_1}{x(x+1)}\frac{K_2}{t(t+1)}=K_1K_2\sum_{x=1}^M\frac{1}{x(x+1)}\sum_{t=1}^T \frac{1}{t(t+1)}\approx K_1K_2.$$
If, for example, we assume that an average client makes a deposit once in a quarter, then $K_1K_2\approx 0.01$ (e.g. $K_1=K_2=0.1$).
Returning to the desired expectation $\eqref{*}$, we get

$$\sum_{k=1}^T\sum_{x=1}^M x \sum_{t=k}^T p(x,t) = 0.01\sum_{k=1}^T\sum_{x=1}^M x \sum_{t=k}^T \frac{1}{x(x+1)t(t+1)}=0.01\sum_{k=1}^T\sum_{x=1}^M \frac{1}{x+1} \sum_{t=k}^T \frac{1}{t(t+1)}=$$

$$0.01\sum_{k=1}^T\sum_{x=1}^M \frac{1}{x+1} \left(\frac{1}{k}-\frac{1}{T+1}\right)=0.01\sum_{k=1}^T\left(\frac{1}{k}-\frac{1}{T+1}\right)\sum_{x=1}^{10^6} \frac{1}{x+1}\approx$$

$$0.01\sum_{k=1}^{10^3}\left(\frac{1}{k}-\frac{1}{10^3}\right)(14.393-1)\approx0.01\cdot 13.393\cdot (7.484-1)\approx 0.87$$
coins in balance from one client on average. 

If we want, for example, 1 million coins in our account on average, then we need about $N\approx1.15$ million clients ($0.87N\approx10^6$), which is fully realistic in our online and mobile world. 

The number of clients depends on the probability of deposit linearly, which means that if the probability of investment is 10 times less ($K_1K_2=0.001$), then we need 11.5 million  clients, which is difficult, but still not impossible to achieve. 

\section{Naive examples of business models}

\subsection{Large software}

Huge expensive software like that of Adobe, Autodesk, SAS, Oracle, Microsoft, Cisco, VxWorks, Maya or Unreal Engine  products needed for high-tech startups are not easy to buy. Startup investors might be happy to use the proposed model to equip their fledgelings with the best technologies at little or no risk.

\subsection{E-books}

Do you really need to buy a book forever \cite{book}? How many books do you read twice? Probably just a few. Then you would feel fair about paying a bit more than usual, like 100-500\$, but getting all your money back in 10-100 days.

\subsection{Games}

The same applies to computer games industry as games lifecycle is shortening (\cite{3,4}).
The range of use of the proposed approach here is practically infinite -- there is a great number of opportunities to improve your character or game process or game world by different temporary features. Which, of course, are given just in exchange for your deposit. It can make a small revolution in games monetization (as it does not seem to be used \cite{5} yet).

\subsection{Membership and service subscription}

Different kinds of membership like closed clubs or service subscriptions are good candidates for the proposed business model. Note that if you do not withdraw your deposit and continue to use it for e.g. playing mobile games or reading books one by one, then the proposed model naturally becomes a usual subscription.

\subsection{Charity foundations}

Also.

\section{What now?}
Probably there are many other directions where ``fair monetization'' may be applied. The author is open for collaboration with institutions and companies interested in development of the described approach.


\begin{thebibliography}{9}

\bibitem{1}
Report: 0.19\% of Players Account for 48\% of Free-to-Play Game Revenue.
``http://www.adweek.com/socialtimes/report-0-19-of-players-account-for-48 -of-free-to-play-game-revenue/636502''

\bibitem{2}
Only 0.15 percent of mobile gamers account for 50 percent of all in-game revenue (exclusive).
``http://venturebeat.com/2014/02/26/only-0-15-of-mobile -gamers-account-for-50-percent-of-all-in-game-revenue-exclusive/''

\bibitem{book}
Reading non-stop, how long does it take you to read a 300 page book?
``https://www.goodreads.com/poll/show/48965-reading-non-stop-how -long-does-it-take-you-to-read-a-300-page-book''

\bibitem{3}
Mobile games lifecycle shortening - App Annie.\\
``http://www.gamesindustry.biz/articles/2016-01-20-mobile-games-lifecycle-shortening -app-annie''

\bibitem{4}
Rise and fall: The numbers behind the lifecycle of mobile games. ``http://www.pocketgamer.biz/comment-and-opinion/60228/the-numbers-behind -the-lifecycle-of-mobile-games/''

\bibitem{5}
42 Strategies for Monetizing Mobile Games.
``http://blog.soom.la/2013/07/42 -ways-to-monetize-your-mobile-game.html''





\end{thebibliography}
\end{document}